\documentclass[journal,onecolumn]{IEEEtran}
\usepackage{cite}

\ifCLASSINFOpdf
  \usepackage[pdftex]{graphicx}
  \DeclareGraphicsExtensions{.pdf,.jpeg,.png}
\else
  \usepackage[dvips]{graphicx}
\fi
\usepackage{amsmath}
\usepackage{amsfonts}

\ifCLASSOPTIONcompsoc
  \usepackage[caption=false,font=normalsize,labelfont=sf,textfont=sf]{subfig}
\else
  \usepackage[caption=false,font=footnotesize]{subfig}
\fi

\newcommand{\z}{\mathbf{z}}
\newcommand{\x}{\mathbf{x}}

\renewcommand{\t}{\mathbf{t}}

\newcommand{\boldxi}{\mbox{\boldmath{$\xi$}}}
\newcommand{\boldalpha}{\mbox{\boldmath{$\alpha$}}}
\newcommand{\boldbeta}{\mbox{\boldmath{$\beta$}}}

\newcommand{\Pbkg}{P_{\mbox{\scriptsize bkg}}}

\newcommand{\Dc}{D_{\mbox{\scriptsize c}}}
\newcommand{\Dglrt}{D_{\mbox{\scriptsize GLRT}}}
\newcommand{\Dbayes}{D_{\mbox{\scriptsize Bayes}}}
\newcommand{\Dlmp}{D_{\mbox{\scriptsize LMP}}}
\newcommand{\Dmixed}{D_{\mbox{\scriptsize mixed}}}

\newcommand{\ie}{\textit{i.e.}}
\newcommand{\eg}{\textit{e.g.}}
\newcommand{\eq}[1]{Eq.~(\ref{eq:#1})}

\usepackage{xcolor}

\begin{document}
%
\title{Homeopathic priors?}
%
%
%

\author{James Theiler\\Los Alamos National Laboratory}

%
%

\markboth{December 5, 2022}{}
%



\maketitle

\begin{abstract}

  The problem of composite hypothesis testing is considered in the
  context of Bayesian detection of weak target signals in
  cluttered backgrounds.  (A specific example is the detection of
  sub-pixel targets in multispectral imagery.)  In
  this model, the target strength (call it $a$) is an unknown parameter,
  and that lack of knowledge can be addressed by
  incorporating a prior over possible parameter values.  The
  performance of the detector depends on the choice of prior, and
  -- with the motivation of enabling better performance at low
  target abundances -- a
  family of priors are investigated in which singular weight is
  associated with the $a\to 0$ limit. Careful treatment of this
  limiting process leads to a situation in which components of the
  prior with infinitesimal weight have
  nontrivial effects.  Similar claims have been made for homeopathic
  medicines.
  
\end{abstract}


%
\IEEEpeerreviewmaketitle

\section{Introduction}
\label{sect:intro}
%
%
%
%


\IEEEPARstart{C}{onsider} the problem of detecting a target of known
spectral signature $\t$, but unknown strength $a$, in a cluttered background.
In this formulation of the problem, the background distribution is known;
specifically, we can write $\Pbkg(\x)$ as the likelihood of observing $\x$
in the background. Depending on how the target interacts with the background,
we can write $p(a,\x)$ as the likelihood of observing $\x$ when the target
strength is $a$.

To be more explicit, write
$\x=\boldxi(a,\z)$ as the effect of a
target with strength $a$ on a background pixel $\z$.  So, for instance,
$\boldxi(a,\z)=\z+a\t$ is the additive model: the effect of a target of
strength~$a$
is to add~$a\t$ to the background~$\z$.  Another common model
arises in spectral imagery, and corresponds to opaque sub-pixel targets
whose strength $a\in[0,1]$ is the fractional area of the pixel that is
occupied by the target\cite[\S11.1.4]{Manolakis16}.  Here $\boldxi(a,\z) =
(1-a)\z + a\t$. Yet another model, associated with gas-phase plume
detection, corresponds to Beer's Law absorption;
here $\boldxi(a,\z)=\z\exp(-a\t)$.  For all of 
these models, we can write
\begin{align}
  p(a,\x) &= \left|\frac{d\boldxi}{d\x}\right|^{-1}
  p(0,\boldxi^{-1}(a,\x)) \nonumber \\
  &= \left|\frac{d\boldxi}{d\x}\right|^{-1}
  \Pbkg(\boldxi^{-1}(a,\x))
\end{align}
where $d\boldxi/d\x$ is the Jacobian, $|\cdot|$ indicates the matrix
determinant, and
$\boldxi^{-1}(a,\x)$ is the value of $\z$ for which $\x=\boldxi(a,\z)$.
For each of these three models, this leads to:
\begin{align}
  \mbox{Additive:~~} & p(a,\x) = \Pbkg(\x-a\t) \\
  \mbox{Replacement\cite{Schaum97,Theiler18igarss}:~~} & p(a,\x) =  (1-a)^{-d}\,\Pbkg\left(\frac{\x-a\t}{1-a}\right) \\
  \mbox{Beer's Law\cite{Theiler20igarss}:~~} & p(a,\x) = \exp(a\tau)\,\Pbkg\left(\x\exp(a\t)\right)
\end{align}
where $d$ is number of spectral channels (\ie,
the dimension of the vectors $\x$ and $\t$), and
$\tau=\sum_\lambda t_\lambda$ is the sum of the components in the vector $\t$.

The null hypothesis, that no target is present, corresponds to
$a=0$. In a conceptually important special case,
the target strength $a=a_o$
\emph{is} known.  That is to say: we know that the target strength is
either $a=a_o$ or $a=0$. This case is one of simple hypothesis
testing, and the optimal detector is given by a likelihood ratio test.
\begin{equation}
  \Dc(a_o,\x) = \frac{p(a_o,\x)}{\Pbkg(\x)} = \frac{p(0,\x-a_o\t)}{p(0,\x)}
  = \frac{\Pbkg(\x-a_o\t)}{\Pbkg(\x)}
  \label{eq:clairvoyant}
\end{equation}
This special case, in which the presence or absence of a target is unknown,
but the target \emph{strength} is known, is unusual in practice, and the
detector has a correspondingly unusual name:  $\Dc(a_o,\x)$ is the \emph{clairvoyant} detector\cite{Kay98}.

A particular clairvoyant detector of interest is what Kay\cite{Kay98}
calls the LMP (locally most powerful) detector; it is given by the
$a\to 0$ limit.  If we explicitly take this limit, however,
we get a null result:
\begin{equation}
  \lim_{a\to 0} \Dc(a,\x) = \frac{\lim_{a\to 0}p(a,\x)}{p(0,\x)} = 1,
\end{equation}
which is not useful, as it does not even depend on $\x$.
The way this is dealt with is to recognize that
any monotonic transform of a detector produces an equivalent detector, and for
any fixed $a>0$, the detector $D^*(a,\x) = (D(a,\x)-1)/a$ is a monotonic
transform of $\Dc(a,\x)$. Thus
\begin{equation}
  \Dlmp(\x) = \lim_{a\to 0} \Dc^*(a,\x) = \lim_{a\to 0}\frac{\Dc(a,\x)-1}{a}
  = \lim_{a\to 0}\frac{d}{da}\Dc(a,\x) = \frac{p'(0,\x)}{p(0,\x)}
\end{equation}
where $p'(0,\x) = \left.\frac{d}{da}p(a,\x)\right|_{a=0}$.  This is
the LMP detector. In general, it \emph{does} depend on $\x$, and in
practice it is a decent detector, though often not optimal because the
$a$'s of interest are often substantially nonzero (especially in
scenarios where low false alarm rates are required).

In the more general case, the null hypothesis is still given by $a=0$,
but the alternative is not a single value $a=a_o$, but \emph{all}
nonzero values: $a\neq 0$.  (Sometimes, the alternative of interest is
given by $a>0$.)  Because the alternative hypothesis involves multiple
values of $a$, this is referred to as the \emph{composite} hypothesis
testing problem. The notion of optimality in this case is a little
ambiguous. In rare cases\footnote{The additive target on a Gaussian
  background is one of those rare cases. Here, the adaptive matched filter\cite{Reed74,Robey92} is the uniformly most powerful detector.}, one can identify a
uniformly most powerful (UMP) detector; a single detector that is
better than every alternative at \emph{every} value of $a$.  When no
UMP detector exists, we can still look for detectors that are
\emph{admissible}.  A detector is admissible if there is no other
detector that is uniformly better. Put another way: given any other
detector, there is some value of $a$ for which the admissible detector
is better.

In the composite hypothesis testing problem, we don't know $a$, so
we end up ``fusing'' the clairvoyant detectors using various strategies. The
classic is the generalized likelihood ratio test (GLRT).
Here, we define $\widehat a(\x)$ as the maximum-likelihood estimate of $a$,
given a measurement $\x$.  This value of $a$ is used as the target
strength in the clairvoyant detector in \eq{clairvoyant}. That leads to:
\begin{equation}
  \Dglrt(\x) = \Dc(\widehat{a}(\x),\x) = \frac{\mbox{max}_a p(a,\x)}{p(0,\x)}
\end{equation}
A generalization of this approach is to use a penalized likelihood
estimator\cite{Chen98,Vexler10}:
\begin{equation}
  \widehat a_q(\x) = \mbox{argmax}_a q(a)\,p(a,\x)
\end{equation}
where the penalty function $q(a)$ can be used to create a detector
with properties that may be preferable to the application at hand.
This idea has been further developed and explored as ``clairvoyant
fusion''\cite{Schaum10,Theiler12spie,Schaum16b}.  It is possible, however,
for GLRT and other clairvoyant fusion detectors to be \emph{inadmissible};
an example was shown in\cite{Theiler12spie}.

Another strategy is to fuse the clairvoyants by averaging them with a
Bayesian prior.  An important advantage to this approach is that
Bayesian detectors are guaranteed to be
admissible\cite{Wald39,Lehmann05}.  The prior is a probability density
$q(a)$ that is traditionally interpreted as the user's prior (\ie,
prior to seeing the data) sense of what the target strength might be.
The Bayesian detector is given by
\begin{equation}
  \Dbayes(\x) = \frac{\int q(a)p(a,\x)\,da}{p(0,\x)}
  \label{eq:bayes}
\end{equation}
In general, one imposes $\int q(a)\,da=1$, and I will do so here.  But
it's not really necessary, since it only affects the overall scale of
$\Dbayes$, and that will not change its performance.  Sometimes one
can even get away with an ``improper'' prior, for which $\int
q(a)\,da=\infty$, but that's not the issue here.
An informal (and not necessarily traditional)
interpretation\footnote{Lehmann and Romano\cite[\S1.6]{Lehmann05} note
  that the assumption that $a$ is a random variable with a known
  distribution ``is usually not warranted in applications.'' Instead,
  they suggest the prior should be interpreted as something that
  ``expresses the importance the experimenter attaches to the various
  values of'' the parameter $a$. Wald\cite[\S3]{Wald39} is even more
  dismissive of interpreting the prior as a probablity distribution
  over a random variable; he writes: ``The reason why we introduce here a
  hypothetical probability distribution of [the unknown parameter] is
  simply that it proves to be useful in deducing certain theorems and
  in the calculation of the best [detector].''}  of $q(a)$ is that it
quantifies the emphasis on getting good performance when the target
strength is $a$.

Observe that when $q(a)=\delta(a-a_o)$ is a ``delta function,'' then
the Bayes detector becomes the clairvoyant detector at $a=a_o$. Using
priors composed from a sum of delta functions\cite{Theiler21igarss}
(or even a single delta function\cite{Theiler21veritas}) is a
particularly convenient choice, because the Bayesian integral in
\eq{bayes} becomes a sum.

\section{Emphasize small target performance by considering the $a\to 0$ limit}

Suppose now that we want to build a detector that puts some emphasis
on the $a\to 0$ performance; \ie, one that is somewhat like the LMP,
but that doesn't put \emph{all} of its emphasis on $a\to 0$.  Just to
be specific, let $q(a) = \alpha q_o(a) + (1-\alpha)q_1(a)$, where
$q_o(a)=\delta(a)$ and $q_1(a)$ is a more traditional prior (\eg, take
$q(a)=1$ just to have a concrete example).  Then we can write
\begin{equation}
  D(\x) = \frac{\int q(a)p(a,\x)\,da}{p(0,\x)}
  = \alpha \frac{\int q_o(a)p(a,\x)\,da}{p(0,\x)}
  + (1-\alpha)\frac{\int q_1(a)p(a,\x)\,da}{p(0,\x)} =
  \alpha + (1-\alpha)\frac{\int q_1(a)p(a,\x)\,da}{p(0,\x)}
\end{equation}
which is, by monotonic transformation, equivalent to the Bayesian detector with
prior $q_1(a)$.  That is: the LMP component is effectively ignored.  The $a=0$ endpoint here seems to be problematic.

\subsection{More formal treatment of the delta function}

If you don't like delta functions, or if you are just a little bit
suspicious of them, consider instead a well-defined exponential:
\begin{equation}
  q_\epsilon(a) = (1/\epsilon)\,e^{-a/\epsilon}
\end{equation}
Observe that $\int q_\epsilon(a)\,da=1$.  We can re-derive
the LMP using this
instead of taking $\delta(a-a_o)$ in the limit as $a_o\to 0$.  Here,
\begin{equation}
  D(\epsilon,\x) = \frac{\int q_\epsilon(a)p(a,\x)\,da}{p(0,\x)}
\end{equation}
If we write out a Taylor series expansion:
\begin{equation}
  p(a,\x) = p(0,\x) + ap'(0,\x) + \frac{1}{2}a^2p''(0,\x) + O(a^3)
\end{equation}
Then we can do the integral\footnote{Actually, if the integral is from 0
  to 1, there is an extra term of $O(e^{-1/\epsilon})$, which we can safely
  neglect for small $\epsilon$.} since we know $\int a^k e^{-a/\epsilon}\,da = k!\, \epsilon^{k+1}$:
\begin{equation}
  \int q_\epsilon(a)p(a,\x)\,da =
  p(0,\x) + \epsilon p'(0,\x) + \epsilon^2 p''(0,\x) + O(\epsilon^3)
\end{equation}
and so
\begin{equation}
  D(\epsilon,\x) = \frac{\int q_\epsilon(a)p(a,\x)\,da}{p(0,\x)} =
  1 + \epsilon\, \frac{p'(0,\x)}{p(0,\x)} + \epsilon^2\, \frac{p''(0,\x)}{p(0,x)} + O(\epsilon^3)
\end{equation}
and for any $\epsilon>0$, we have $D^*(\epsilon,\x) =
(D(\epsilon,\x)-1)/\epsilon$ is just a monotonic rescaling, so
\begin{equation}
  \Dlmp(\x) = \lim_{\epsilon\to 0} D^*(\epsilon,\x) = \frac{p'(0,\x)}{p(0,\x)}
\end{equation}
which is the same result as obtained previously, but without invoking
delta functions.

The problem is: when we apply this to the mixed prior, to $q(a) =
\alpha q_\epsilon(a) + (1-\alpha)q_1(a)$, it doesn't seem to help.
The monotonic scaling that saved the integral in the first term blows
up the second term.
\begin{align}
  \Dmixed(\epsilon; \x) &= \frac{\int q(a)p(a,\x)\,da}{p(0,\x)} =
  \alpha\,\frac{\int q_\epsilon(a)p(a,\x)\,da}{p(0,\x)} +
  (1-\alpha)\,\frac{\int q_1(a)p(a,\x)\,da}{p(0,\x)} \nonumber \\
  &=
  \alpha + \alpha\epsilon \frac{p'(0,\x)}{p(0,\x)} + O(\epsilon^2)
  +(1-\alpha)D_1(\x) \nonumber \\
  &=
  \alpha + \alpha\epsilon \Dlmp(\x) + (1-\alpha)D_1(\x) + O(\epsilon^2)
  \label{eq:Dmixed}
\end{align}
where $D_1(\x)$ is the Bayesian detector with the smooth (\eg,
uniform) prior $q_1(a)$.  For any monotonic rescaling that we could
come up with, the $D_1$ term will dominate the $\Dlmp$ term as
$\epsilon\to 0$.

\subsection{More careful treatment of the mixture parameter}

The idea here is still to use \eq{Dmixed} but to have $\alpha\to 1$ so
that the $D_1$ term does \emph{not} overwhelm the $\Dlmp$ term.  Let
\begin{equation}
  \beta = \frac{\epsilon}{1-\alpha+\epsilon}
\end{equation}
so then $\alpha = 1 + \epsilon - \epsilon/\beta$.  If we think of
$\beta$ as the quantity that is fixed as $\epsilon\to 0$, then
$\alpha$ is \emph{not} fixed, and
\begin{equation}
  \Dmixed^*(\epsilon; \x) = \frac{\beta(\Dmixed(\epsilon; \x) -
    \alpha)}{\epsilon} = \beta\alpha\Dlmp(\x) + (1-\beta) D_1(\x) +
  O(\epsilon)
\end{equation}
is a monotonic transformation of $\Dmixed$ in \eq{Dmixed}.  Further,
it has a very well-defined $\epsilon\to 0$ limit, given by
\begin{equation}
  \Dmixed^*(\x) = \lim_{\epsilon\to 0} \Dmixed^*(\epsilon; \x)
  = \beta\Dlmp(\x) + (1-\beta) D_1(\x),
  \label{eq:Dmixed-star}
\end{equation}
which enables us to mix $\Dlmp(\x)$ and $D_1(\x)$ with adjustable
proportions. This is precisely what we set out to do: to incorporate
an adjustable amount of $\Dlmp$ into our detector design.

But the prior associated with this mixture is:
\begin{align}
  q(\epsilon,a) &= \alpha q_\epsilon(a) + (1-\alpha) q_1(a) \nonumber \\
  &= \left(1+\epsilon-\epsilon/\beta\right)q_\epsilon(a)
  + \epsilon(1/\beta-1) q_1(a) \label{eq:prior-vs-e} \\
  &\to \delta(a) \mbox{~as~} \epsilon\to 0.
\end{align}
The infinitesimal $\epsilon(1/\beta-1) q_1(a)$ component of the prior
produces a decidedly finite component $(1-\beta)D_1(\x)$ in the
detector that is derived from that prior.

\begin{figure}
  \centerline{
    \parbox[c]{0.45\textwidth}{
      \includegraphics[width=0.5\textwidth]{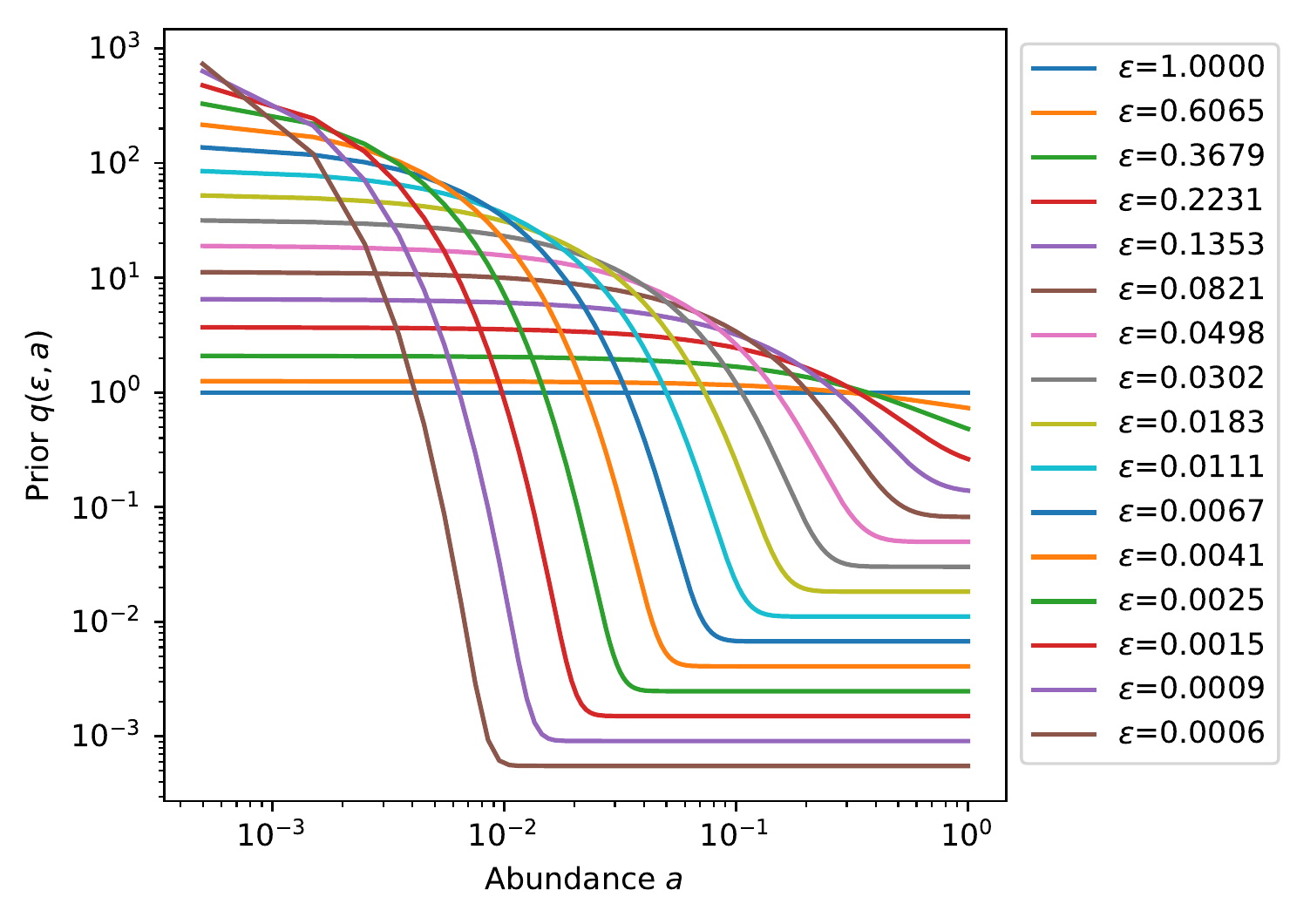}}\hfill
  \parbox[c]{0.45\textwidth}{\caption{Log-log plot of the mixed prior
      $q(\epsilon,a)$, as defined in \eq{prior-vs-e}, as a function of
      the abundance~$a$, for various values of $\epsilon$. As
      $\epsilon\to 0$, the prior becomes more and more like a delta
      function at $a=0$.  In particular, the \emph{value} of
      $q(\epsilon,a)\to 0$ as
      $\epsilon\to 0$ for all $a>0$.  Despite this, the \emph{effect} of
      $q(\epsilon,a)$ for $a>0$ is tangible even as $\epsilon\to
      0$.}}}
  \label{fig-prior-vs-e}
\end{figure}

\section{Further Remarks}

Much as I would like to end on the homeopathic note, where
infinitesimal components in the prior can have a tangible effect on
the resulting detector, there is a different interpretation.  Lehmann
and Romano\cite[\S1.8]{Lehmann05} note that the set of admissible
detectors includes all Bayesian detectors and all \emph{limits} of
Bayesian detectors.  The mixed detector in \eq{Dmixed-star} is one
such limit.  It is admissible, and it is the limit of Bayesian
detectors, but it is not itself a Bayesian detector, because there is
no prior that can be used to derive it (except in the limiting way
that was done here).

This may have consequences for the ``sculpting'' of priors.  Although
the traditional interpretation of a prior is a probability density
that describes the parameter~$a$, that interpretation is not very
useful to designing detectors.  However, it may make sense to choose
priors that optimize specified properties of detectors.  In the course
of optimizing priors, however, one should keep in mind that an
admissible (and possibly optimal in the context of some desired
criteria) detector may not have a well-defined prior associated with
it.

More specifically, a convenient way to sculpt a prior is to
parameterize it with a sum of delta functions, but with special
treatment for $a=0$.  In particular,
\begin{equation}
  q(\boldalpha,a) = \alpha_o\delta(a) + \sum_i \alpha_i\delta(a-a_i)
\end{equation}
is associated with
\begin{equation}
  D^*(\boldbeta,\x) = \beta_o\Dlmp(\x) + \sum_i \beta_i D_c(a_i,\x),
\end{equation}
where $\alpha_o = 1 - \epsilon(1-\beta_o)/\beta_o + O(\epsilon^2)$ and
$\alpha_i=\epsilon\beta_i/\beta_o + O(\epsilon^2)$.  Since we care
about the small $\epsilon$ limit, we are better off trying to optimize
the $\boldbeta$ coefficients instead of trying to directly estimate
the best prior by optimizing the $\boldalpha$ coefficients.

\bibliography{veritas}
\bibliographystyle{IEEEtran}

\end{document}